\documentclass[preprint, 12pt, dvipsnames]{elsarticle}

\usepackage{amssymb}
\usepackage{braket}
\usepackage{bm}
\usepackage{mathtools}
\usepackage{float}
\usepackage[colorlinks]{hyperref}

\usepackage{tikz}
\usetikzlibrary{positioning}
\usepackage{soul}

\usepackage{amsthm}
\usepackage{algorithm2e}

\newcommand{\nrgerr}{\mathcal{E}_r} 
\newcommand{\lossover}{\mathcal{L}_{\rm Overlap}} 
\newcommand{\lossmse}{\mathcal{L}_{\rm MSE}}

\affiliation[1]{organization={Science, Mathematics and Technology Cluster, Singapore
University of Technology and Design},
addressline={8 Somapah Road},
postcode={487372},
city={Singapore},
country={Singapore}}

\affiliation[2]{organization={Max Planck Institute for the Science of Light},
addressline={Staudtstraße 2},
postcode={91058},
city={Erlangen},
country={Germany}}

\affiliation[4]{organization={Engineering Product Development Pillar},
addressline={Singapore University of Technology and Design},
postcode={487372},
city={Singapore},
country={Singapore}}

\affiliation[5]{organization={Centre for Quantum Technologies},
addressline={National University of Singapore, 3 Science Drive 2},
postcode={117543},
city={Singapore},
country={Singapore}}

\affiliation[6]{organization={MajuLab, CNRS-UCA-SU-NUS-NTU International Joint Research Unit},
addressline={Radarweg 29},
postcode={1043 NX},
city={Singapore},
country={Singapore}}

\affiliation[3]{organization={School of Computing, National University of Singapore},
addressline={13 Computing Drive},
postcode={117417},
city={Singapore},
country={Singapore}}

\journal{Computer Physics Communications}

\begin{document}

\title{Supervised Training of Neural-Network Quantum States for the Next-Nearest Neighbor Ising Model}

\author[1]{Zheyu Wu
\fnref{thefootnote}}

\author[2]{Remmy Zen
\fnref{thefootnote}}

\author[1]{Heitor P. Casagrande}

\author[1,4,5,6]{Dario Poletti 
\fnref{thefootnote2}}

\author[3]{St\'ephane Bressan}

\fntext[thefootnote]{These authors contributed equally to this work.}

\fntext[thefootnote2]{dario\_poletti@sutd.edu.sg to whom corresponcence should be sent}

\begin{abstract}

Different neural network architectures can be unsupervisedly or supervisedly trained to represent quantum states. We explore and compare different strategies for the supervised training of feed forward neural network quantum states. 
We empirically and comparatively evaluate the performance of feed forward neural network quantum states in different phases of matter for variants of the architecture, for different hyper-parameters, and for two different loss functions, to which we refer as \emph{mean-squared error} and \emph{overlap}, respectively.
We consider the next-nearest neighbor Ising model for the diversity of its phases and focus on its paramagnetic, ferromagnetic, and pair-antiferromagnetic phases. 
We observe that the overlap loss function allows better training of the model across all phases, provided a rescaling of the neural network. 
\end{abstract}

\maketitle

\section{Introduction}
\label{sec:introduction}

Machine learning models, in general ~\cite{CarleoZdeborova2019, MelkoCirac2019, CarrasquillaTorlai2021, biamonte2017quantum, DunjkoBriegel2018, MehtaSchwab2019}, and neural networks, in particular, have been used in a variety of applications including phase characterization~\cite{Wang2016, carrasquilla2017machine, vanNieuwenburgHuber2017}, experimental guidance \cite{WeiLei2019}, state tomography~\cite{Lanyon2017, Torlai2018, XuXu2019, Carrasquilla2019, Westerhout2020, QuekNg2021, Koutny2022}, process tomography \cite{Banchi2018, GuoPoletti2020}, control~\cite{Bukov2018, Bukov2018b, JerbiDunjko2021}, the learning of Hamiltonians~\cite{BurgarthNori2009, BaireyNetanel2019, Ma2021, WildeEisert2022, XiaoZeng2022} and dissipators~\cite{BaireyArad2020, VolokitinDenisov2022}, and the learning of wave functions for both equilibrium and non-equilibrium cases~\cite{carleo2017solving, HartmannCarleo2019, VicentiniCiuti2019, YoshiokaHamazaki2019, NagySavona2019, schmitt2020quantum, GutierrezMendl2022, Nomura2021, ParkKastoriano2020, ParkKastoriano2021, ChooNeupert2018, ZenBressan2020, ZenBressan2020b, RothMacDonald2021, kochkov2018variational, VerdelHeyl2021,DonatellaCiuti2022,JonssonCarleo2018, MedvidovicCarleo2021, SinibaldiVicentini2023}. 

In the latter application, \emph{neural-network quantum states}, originally proposed in Ref. \cite{carleo2017solving}, refer to a family of neural networks trained to approximate the wave function of a quantum system, namely to return unnormalized probability or the probability amplitude corresponding to each configuration of the quantum system. 

Numerical methods to study quantum systems face the challenge that the Hilbert space size increases exponentially with the number of components. It is not possible for any method on a classical computer to describe all the possible wave functions. However, one can search for methods to accurately describe physically relevant states. In this work, we focus on ground states of the one-dimensional next-nearest neighbor quantum Ising model because the ground state can be in four different phases: paramagnetic, ferromagnetic, antiferromagnetic, and pair-antiferromagnetic phases. 
While the classical Ising model has already been investigated with the use of neural networks \cite{KimKim2018, EfthymiouMelko2019, DAngeloBottcher2020}, the quantum Ising model with next-nearest neighbor couplings is more complex and requires deeper investigations. 
An important point in the use of neural network is its trainability. Even if a network could represent a certain wave function or probability distribution, it may not be effective or efficient to train it to reach that goal function. Here, in particular, we wonder, within the same structure of the neural network, which optimization routines can allow us to describe accurately the very different phases of these ground states.


Neural-network quantum states can be unsupervisedly or supervisedly trained. Unsupervised neural-network quantum states, for instance, variationally optimize the model parameters to minimize the energy of a system~\cite{carleo2017solving}. Unsupervised learning is frequently performed by stochastic reconfiguration~\cite{sorella2007weak}, and can be accelerated by transfer learning~\cite{ZenBressan2020, ZenBressan2020b}, sequential local optimization~\cite{ZhangPoletti2023} or by computing the geometric tensor on the fly~\cite{VicentiniCarleo2021}. 

Supervised neural-network quantum states are trained with a (large) corpus of configurations and their probabilities or probability amplitudes~\cite{carrasquilla2017machine} obtained from other numerical simulations or, more importantly, from experiments. Supervised learning has two main applications. Firstly, it reveals whether a particular model is expressive enough and sufficiently easily trainable to represent a specific wave function in practice. This information can then be used to select that model to describe similar systems or classify the types of physical systems and correlations that the machine learning model can describe. Secondly, it yields a machine representation of the wave function that can further be used to compute additional elements and gather insights into the physical system, which may otherwise be challenging to obtain experimentally. For instance, not all observables (e.g. long-range correlations) and measurables (e.g. entropy and out-of-time-ordered correlators~\cite{LarkinOvchinnikov1969}) are readily accessible on experimental platforms. 

To gain general strategies on how to effectively perform supervised learning, here we explore and compare different strategies for the supervised training of a neural-network quantum states architecture consisting of a feed forward neural network. 
We comparatively evaluate the performance, namely the effectiveness of the training, of feed forward neural network quantum states in different phases of matter for variants of the architecture, for instance, different number of hidden nodes, for different hyper-parameters, different learning rates, and different number of samples. In particular, we consider two different loss functions, to which we refer as \emph{mean-squared error} and \emph{overlap}, respectively. 

As we mentioned before, we consider the one-dimensional next-nearest neighbor quantum Ising model. What is important to add here is that this model also has the advantage that it can be studied very accurately and efficiently using matrix product state algorithms \cite{schollwock2011density}. We will thus use this algorithm to produce the exact ground state from which we will also sample the configurations and thus provide the data to train the neural networks. This will allow us to effectively study the expressive power of the networks and the possibility of training them.

The paper is structured as follows. Section~\ref{sec:sup_nnqs} introduces the neural network architecture, the supervised training, and the two loss functions considered. Section~\ref{sec:model} describes the physical system studied and recalls its properties. Section~\ref{sec:results}  presents and analyzes the results of the comparative empirical evaluation.  Finally, Section~\ref{sec:conclusions} summarizes the findings.

\section{Supervised neural-network quantum states}
\label{sec:sup_nnqs}

\begin{figure}[ht]
\centering
\includegraphics[width=0.6\columnwidth]{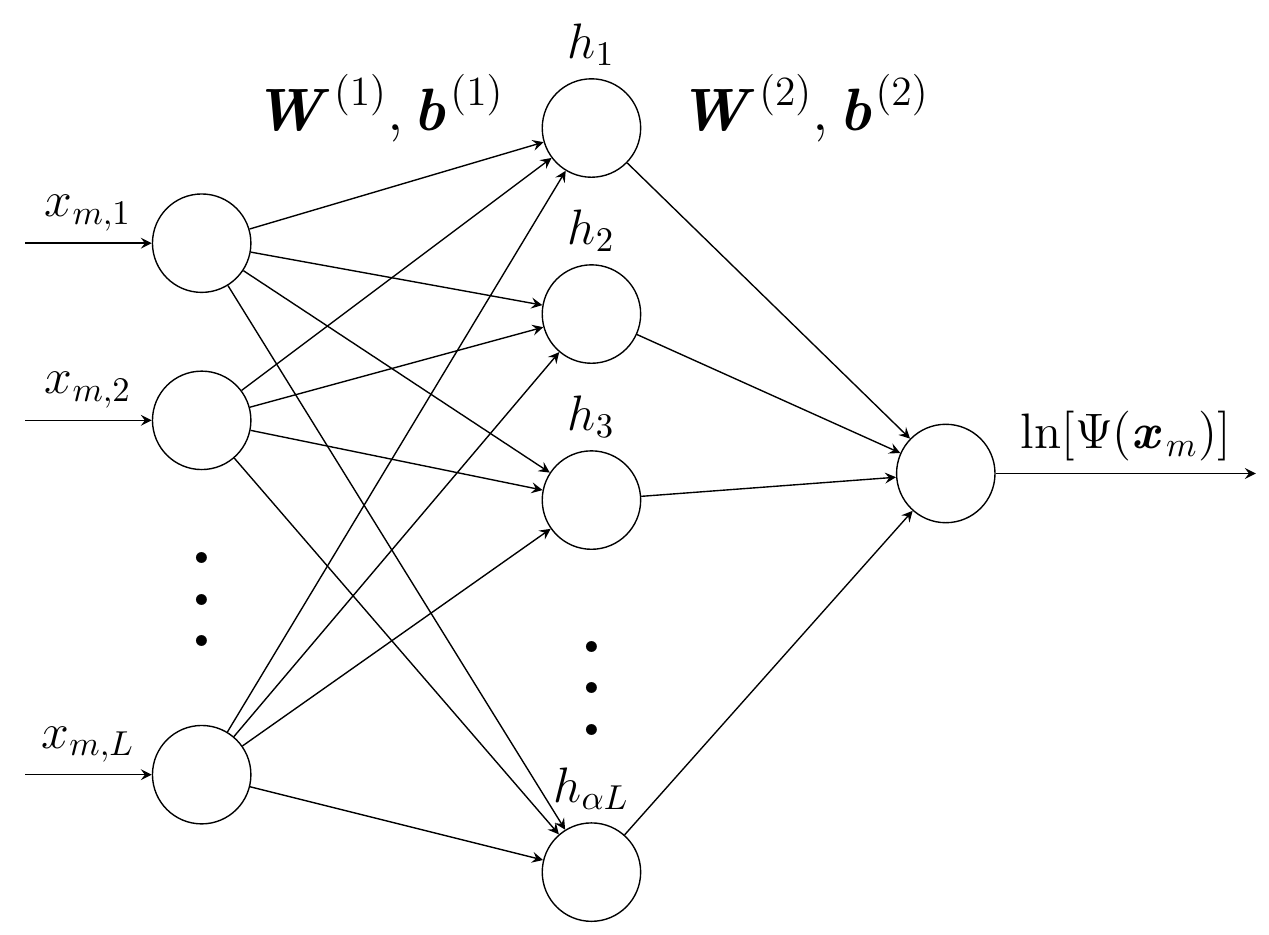}
    \caption{Feed forward neural network architecture. Inputs are $x_{m,i}$, the elements of a spin configuration ${\bm x_m}$, $L$ is the number of spins of the system, with one hidden layer of size $\alpha L$ and output $\ln\left[\Psi({\bm x_m})\right]$. $W$ and $b$ are the weights and biases, respectively.}
    \label{NNarchitecture}
\end{figure}

Neural-network quantum states belong to the family of variational Monte Carlo methods and use neural networks as the trial wave function. They were proposed by the authors of~\cite{carleo2017solving} to represent with a neural network the amplitude $\Psi(\bm{x_m})$ of the configurations $\bm{x_m}$ of a quantum state. 
Here, our choice of the neural network is a feed forward neural network, where each visible node represents one of the $L$ particles of the quantum many-body system. More specifically, a visible node can take the values $-1$ or $1$ depending on whether the spin at that location is pointing down or up, respectively. Figure~\ref{NNarchitecture} depicts a feed forward neural network with a single hidden layer. Note that the output layer returns the natural logarithm of the probability amplitude associated with one configuration, i.e. $\ln[\Psi({\bm x_m})]$, because the wave function typically takes values that vary over different orders of magnitude. One advantage of a feed forward neural network is its flexible design and easier to construct deeper structure.

In feed forward neural-network quantum states, we use a concatenation of linear and nonlinear functions given by 
\begin{equation}
\label{eq:mlpnqs}
 \ln\left[\Psi(\bm{x_m})\right]  =  \sigma^{(k)}(\bm{W}^{(k)} \dots  \sigma^{(1)}(\bm{W}^{(1)}\bm{x_m} + \bm{b}^{(1)})  \dots + \bm{b}^{(l)}),  
\end{equation} 
where $\bm{W}^{(k)}\bm{x_m} + \bm{b}^{(k)}$ is the linear part at layer $k$, $\bm{W}^{(k)}$ and $\bm{b}^{(k)}$ represent the neural network parameters at layer $k$. $\bm{W}^{(k)}$ are the parameters which fully connect the nodes between adjacent layers and $\bm{b}^{(k)}$ are the bias. While $\sigma^{(k)}$ is the nonlinear activation function at layer $k$.
For our purpose, it is sufficient to consider a single hidden layer. The number of nodes in this layer is given by $\alpha L$. We refer to $\alpha$ as the hidden ratio. For the activation function $\sigma^{(k)}$, for $k=1$ we use leaky ReLU~\cite{Maas2013RectifierNI}, which is given by 
\begin{equation}
    \text{LeakyReLU}(x)= 
\begin{cases}
    x                   & \text{if } x\geq 0\\
    \eta x            & \text{otherwise}
\end{cases}
\label{LeakyRelu}
\end{equation}
where $\eta$ is a constant we choose to fix it to 0.01 for our experiments. This activation function can prevent vanishing gradient, which could happen with ReLU performing on negative input.

The supervised training of neural-network quantum states is analogous to the regression problem in standard machine-learning settings. We used labeled data $(\bm{x_m}, \ln[\Psi_G(\bm{x_m})])$ pairs, where $\Psi_G(\bm{x_m})$ is the exact value of probability amplitude for the spin configuration $\bm{x_m}$. Here $m$ goes from $1$ to $M$, while $M_u$ is the number of unique samples, as sometimes many configurations could be repeated. The samples should be representative of the probability distribution stemming from the wave function, and they could come from experimental data, although, here, we use numerically generated ones. 
In the following, we also use the notation $\ln[\tilde{\Psi}({\bm x}_m)]$ for the rescaled output of the neural network, which is given by 
\begin{align}
    \ln[\tilde{\Psi}({\bm x}_m)] = \frac{\ln[\Psi({\bm x}_m)]}{\max_k \ln[\Psi({\bm x}_k)]}, \quad k \in [1,M_u] \label{NNoutput}
\end{align}
which is the output of the neural network divided by the maximum occurring for the configurations considered. From this it follows that 
\begin{equation}
    \tilde{\Psi}({\bm x}_m) = \exp[ \ln [ \tilde{\Psi}({\bm x}_m) ] ]. 
    \label{expLnConc}
\end{equation}

To obtain $\Psi_G(\bm{x_m})$, we use a matrix product state algorithm with the zipper method to obtain the samples~\cite{HanZhang2018, PeresCasagrandePoletti2023, samplingDataGit}. While the total size of the possible labeled data is $2^L$, where $L$ is the size of the system under study, in general, we can only take a much smaller number of sample $M$ drawn from the distribution $|\Psi_G(\bm{x_m})|^{2}$ using the Metropolis-Hastings algorithm~\cite{MetropolisTeller1953, Hastings1970}. 

Among the variety of possible loss functions, we consider minimizing two different possible loss functions to train the neural network: the mean-squared error (MSE) which is commonly used to conduct supervised learning, given by 
\begin{equation}
\label{eq:mse}
\mathcal{L}_{\text{MSE}}(\Psi_G, \Psi) = \frac{1}{M} \sum_{m = 1}^{M} \Big(\ln{[\Psi_G(\bm{x_m})]} - \ln{[\Psi(\bm{x_m})]}\Big)^2, 
\end{equation}
and the overlap between two wave functions, given by 
\begin{equation}
\label{eq:overlap}
\begin{split}
\mathcal{L}_{\text{Overlap}}(\Psi_G, \tilde{\Psi})  &= -\ln \left[ \frac{\braket{\Psi_G \mid \tilde{\Psi}} \braket{\tilde{\Psi} \mid \Psi_G}}{\braket{\Psi_G \mid \Psi_G} \braket{\tilde{\Psi} \mid \tilde{\Psi}}} \right]
\end{split}. 
\end{equation}
We note that, in Equation~(\ref{eq:overlap}), we can rewrite each term according to 
\begin{equation}
    \braket{\Psi_A \mid \Psi_B} = \sum_{m = 1}^{M_u} \Psi_A^*(\bm{x_m})\Psi_B(\bm{x_m}). 
    \label{eq:scalar_product}
\end{equation}
This is also used in Ref.~\cite{cai2018approximating}. Here we used $\tilde{\Psi}_{\beta}(\bm{x_m})$, where $\beta=A, B \dots$, and not $\Psi_{\beta}(\bm{x_m})$ because we need to exponentiate the output of the neural network and this can lead to instabilities, see Ref.~\cite{VicentiniCarleo2021} and~\ref{app:notrescaled}. Note that in Equation~(\ref{eq:overlap}-\ref{eq:scalar_product}), we use the unique samples $M_u$ instead of all the samples. This is because in the calculation of the overlap there is already the probability of each unique samples, and if we were to take all the samples into account (with repetitions proportional to the probability), the probability of the samples would be weighted twice. As a result, the normalization would skewed towards the more probable configurations\footnote{Note that one could use unique samples also for the MSE loss function by slightly modifying it to  
\begin{equation}
\mathcal{L}_{\text{MSE}}(\Psi_G, \Psi) = 
\sum_{m = 1}^{M_u} |\Psi_G({\bm x_m})|^2\Big(\ln{[\Psi_G(\bm{x_m})]} - \ln{[\Psi(\bm{x_m})]}\Big)^2. \label{eq:mse_app}
\end{equation}

However, in our empirical evaluations, we find that the loss function which considers all the samples has a more consistent response. This could be investigated further in future works.}. We then compute the gradient of the loss functions and update the parameters using an initial learning rate $\gamma$ and Adam optimizer introduced in Ref.~\cite{kingma2017adam}.\footnote{In the initialization of the neural network, and generation of the batches, we use the same random seed for different numerical experiments.}


\section{The next-nearest neighbor Ising model}\label{sec:model} 


\begin{figure}[ht]
\centering
\includegraphics[width=\columnwidth]{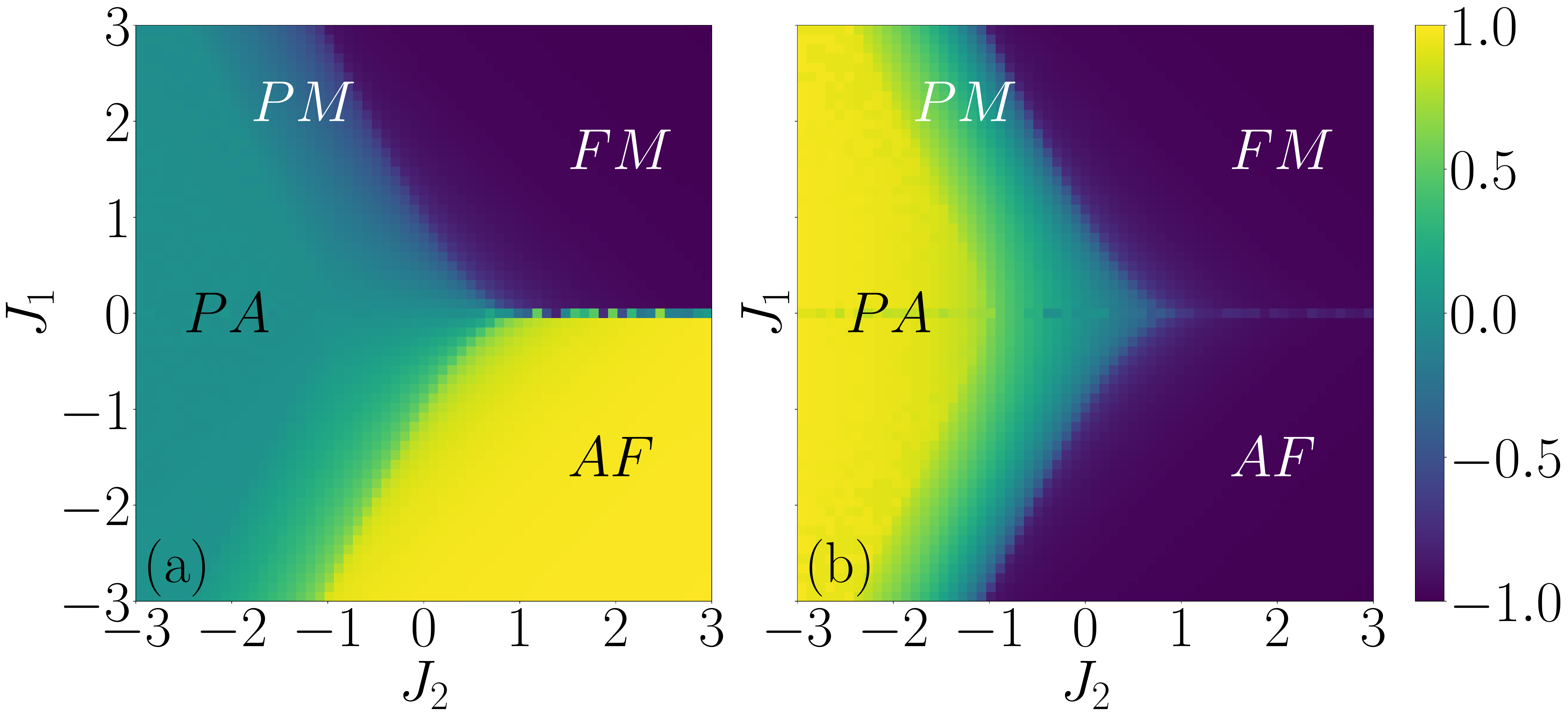}
\caption{Phase diagram of the next-nearest neighbor Ising model, Equation~(\ref{eq:hamiltonian})  with $h_x = 1$ and system size $L=128$,  in terms of tunneling $J_1$ and $J_2$ interactions. (a) Nearest-neighbor correlation $C_{NN}$ from Equation~(\ref{eq:cnn}); (b) Next-nearest neighbor correlation $C_{NNN}$ from Equation~(\ref{eq:cnnn}). Paramagnetic region is labeled by $PM$, ferromagnetic by $FM$, antiferromagnetic by $AF$, and pair-antiferromagnetic by $PA$.}
    \label{fig:PhaseDiagrams}
\end{figure}

While supervised training can, in principle, be used to learn any state, we focus on training ground states with different properties. The main reason why we use ground states is that with the methods we use to generate data, i.e. matrix product states, ground states can be efficiently and effectively produced \cite{Schollwock2011}.  
As for the physical model studied, for our analysis, we choose the next-nearest neighbor Ising model 
because its ground state can be in four different phases.
More specifically, the Hamiltonian is given by
\begin{equation}
    H = -h_x \sum_{l=1}^L\sigma^x_l - J_1 \sum_{l=1}^L \sigma^z_l \sigma^z_{l+1} - J_2 \sum_{l=1}^L \sigma^z_l \sigma^z_{l+2}\label{eq:hamiltonian}, 
\end{equation}
where $\sigma^a_l$ are the corresponding Pauli matrices (with $a=x,y,z$) at site $l$, $h_x$ is the magnetic field along the $x$ axis, and $J_1$ and $J_2$ are nearest neighbor and next-nearest neighbor couplings, respectively. We note that the Hamiltonian has a ``particle-hole'' symmetry such that one can map a Hamiltonian with coupling $J_1$ to one with coupling $-J_1$ via the operator $S = \prod_{j} \sigma^x_{2j}$, i.e. $H(J_1) = SH(-J_1)S^\dagger$.  
This transformation, for instance, turns a ferromagnetic phase into an antiferromagnetic phase. It is thus sufficient to study the system for positive values of $J_1$. The phase diagram, in terms of $J_1$ and $J_2$, is depicted in Figure~\ref{fig:PhaseDiagrams} for $L = 128$ and $h_x=1$, which is the energy scale we use throughout the paper. 
To compute the ground state from which to obtain the training data, we have used a matrix product states code built using \textsc{ITensor} libraries ~\cite{itensor, odmrgLib}. To identify the phases we use the nearest-neighbor ferromagnetic correlator $C_{NN}$ given by 
\begin{align}
    C_{NN} = -\dfrac{1}{L-1} \sum_{l = 1}^{L-1} \sigma^{z}_{l} \sigma^{z}_{l+1}, \label{eq:cnn}
\end{align}
depicted in Figure~\ref{fig:PhaseDiagrams}(a), and the next-nearest neighbor antiferromagnetic correlator $C_{NNN}$, given by 
\begin{align}
    C_{NNN} = \dfrac{1}{L-2} \sum_{l = 1}^{L-2} \sigma^{z}_{l} \sigma^{z}_{l+2}, \label{eq:cnnn}
\end{align}
shown in Figure~\ref{fig:PhaseDiagrams}(b).

In the ferromagnetic phase, $C_{NN}$ and $C_{NNN}$ are both large and negative. 
In the pair-antiferromagnetic phase, e.g. spin configurations of the type $\uparrow\uparrow\downarrow\downarrow\uparrow\uparrow\downarrow\downarrow$, only $C_{NNN}$ is large and positive while $C_{NN}$ approaches zero. The paramagnetic phase is located between these two phases, for which both order parameters are small \footnote{Along the line $J_1 = 0$ we have a pixelated region which is due to the small energy gap and the significant difference in properties of the lowest energy eigenstates. We note that we did not use data for $J_1 = 0$ to train the neural network.}. For negative $J_1$, we can obtain the possible phases by applying the operator $S$, resulting in an antiferromagnetic phase and another pair-antiferromagnetic phase, separated by a paramagnetic phase. We also note that for $J_1>0$, the probability amplitude $\Psi({\bm x_m})$ can be written as a positive real number. Thus one can use a feed forward neural network model with real weights and biases without complex numbers.  

\section{The Mapalus Library}
Carleo et al. presented the first open-source library for neural-network quantum states, called \textsc{NetKet}, in~\cite{carleo2019netket}. Initially, the core of \textsc{NetKet} was implemented in C++ with an interface to Python. However, it is now a Python library that uses the \textsc{Jax} library~\cite{jax2018github} to support general-purpose graphics processing units.


We developed a library, called \textsc{Mapalus}, that implements neural-network quantum states with the \textsc{TensorFlow} library~\cite{tensorflow2015-whitepaper}. We choose \textsc{TensorFlow} over other deep learning frameworks for its widespread use~\cite{zhang2021comparative} and user-friendly application programming interface. \textsc{Mapalus} is available at \url{https://github.com/remmyzen/mapalus} and is the library used in this work. See~\ref{app:mapalus} for more details.

\section{Results}\label{sec:results} 
In this section, we study the performance of training the feed forward neural network. We consider a large system with $L=128$ in three different phases namely the paramagnetic phase, the ferromagnetic phase, and the pair-antifer\-romagnetic phase. We comparatively evaluate the performance for the two different loss functions $\lossover$ and $\lossmse$ together with various training parameters. One important training parameter is the number of training samples $M$, which, if not stated otherwise, we set to $10^5$. For $\lossover$, we first obtain the $M$ samples, then only use the corresponding $M_u$ unique ones to update the neural network. 
During the training, we typically only use a portion of these data to evaluate the loss function and optimize the parameters at each epoch. We call the number of samples in this portion of the data as \emph{batch size} and refer to it with the symbol $B$, which, if not stated otherwise, we set to $B = 32$. Another parameter we vary is the hidden ratio $\alpha$, which, if not stated otherwise, we set to $\alpha=3$.

To evaluate the effectiveness of the training, we study the relative energy error $\nrgerr$ defined by 
\begin{align}
    \nrgerr = \left|\frac{E_T - E_G}{E_G}\right|,  \label{eq:rel_nrg} 
\end{align}
where $E_G$ is the ground state energy computed from matrix product states simulations, while $E_T$ is the energy of the trained wave function. For the trained wave function, we evaluate the energy by  
\begin{align}
    E_T = \frac{1}{M} \sum_{m=1}^M  E_{\rm loc}({\bm x_m}) \label{eq:nrg_all}  
\end{align} 
when using $\lossmse$, as we use all the $M$ samples, and 
\begin{align}
    E_T = \sum_{m=1}^{M_u} |\Psi(\bm x_m)|^2 E_{\rm loc}({\bm x_m}) \label{eq:nrg_unique}  
\end{align} 
when using $\lossover$ because it uses only the unique samples $M_u$. 

In Equations~(\ref{eq:nrg_all}) and (\ref{eq:nrg_unique}), the so-called local energy, local in the sense of configurations, $E_{\rm loc}$ is given by 
\begin{align}
    E_{\rm loc}({\bm x_m}) = \sum_{m'} H_{m',m} \frac{\Psi({\bm x_{m'}})}{\Psi({\bm x_m})}, \label{localEnegy}
\end{align}
where $H_{m',m}$ is the matrix element of the Hamiltonian which connects the two configurations ${\bm x_m}$ and ${\bm x_{m'}}$. Naturally, if we already knew what the Hamiltonian was, we could have done the unsupervised training, but here we use the Hamiltonian to compute a ground truth reference and quantify the effectiveness of the method.

\subsection{Training in the paramagnetic phase}

\begin{figure}[ht]
\centering
\includegraphics[width=0.95\columnwidth]{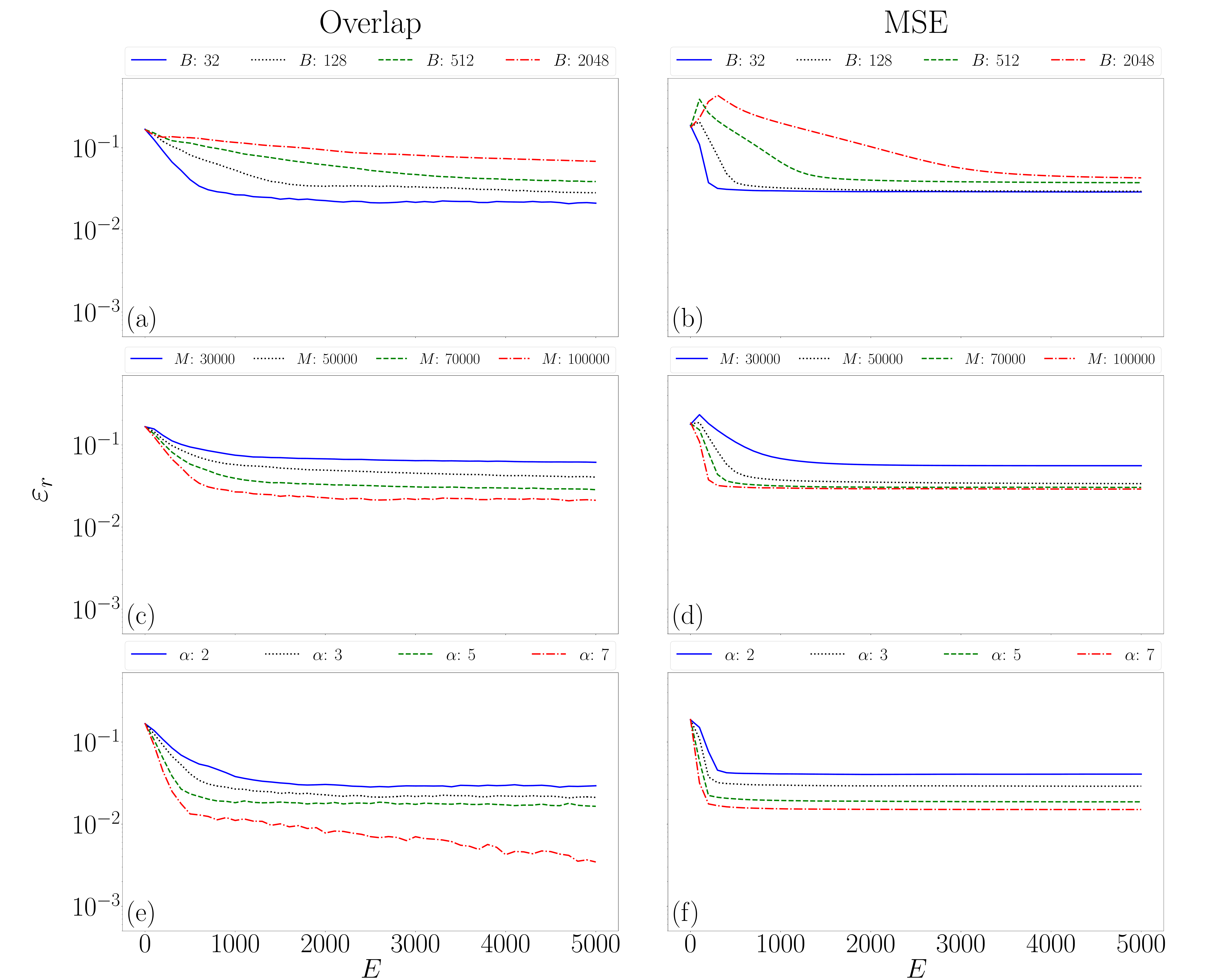} 
    \caption{Relative error $\nrgerr$ for the paramagnetic ground state ($J_1=2$ and $J_2=-1$) versus the number of epochs $E$. In panels (a, c, e), we use overlap loss function $\lossover$; in panels (b, d, f), we use $\lossmse$. In panels (a,b), we vary the batch size $B$. In panels (c,d), we vary the number of samples $M$; in panels (e,f), we vary the hidden ratio $\alpha$. If not stated otherwise, the number of training samples is $M=10^5$, the hidden ratio is $\alpha=3$, the batch size is $B=32$ and $\gamma=10^{-5}$ is the initial learning rate.} 
    \label{fig:para}
\end{figure}

The results for the paramagnetic phase are shown in Figure~\ref{fig:para}. 
In Figure~\ref{fig:para}(a,c,e), we show results for the loss function $\mathcal{L}_{\rm Overlap}$, while in panels (b,d,f) we have considered $\mathcal{L}_{\rm MSE}$. In panels (a) and (b) we keep all parameters as default values mentioned earlier except for the batch size $B$, in panels (c) and (d) we vary the number of samples $M$, and in panels (e) and (f) we consider different hidden ratios $\alpha$. 
Here, the parameters in Hamiltonian are $J_1 = 2$ and $J_2 = -1$, and we have used an initial learning rate $\gamma=10^{-5}$. The default parameters, unless stated otherwise, are the number of samples $M = 10^5$, batch size $B = 32$, and hidden ratio $\alpha=3$. 
 
In all panels, we look at the relative error of the energy $\nrgerr$ versus the number of epochs $E$.   
We observe that both loss functions allow reaching low relative errors of the order of $10^{-2}$.\footnote{When using an initial learning rate $\gamma=10^{-4}$, with $\lossover$ it is possible to reach even better accuracy of the energy, however, the learning dynamics is more involved (not shown). Using $\lossmse$ instead, the prediction of the ground state energy does not improve significantly.} 
In general, both a smaller batch size (see~\ref{app:batch}), and a larger number of samples result in better accuracy. Furthermore, for both loss functions, a larger hidden ratio can also result in better accuracy.


\subsection{Training in the ferromagnetic phase}

\begin{figure}[ht]
\centering
\includegraphics[width=0.95\columnwidth]{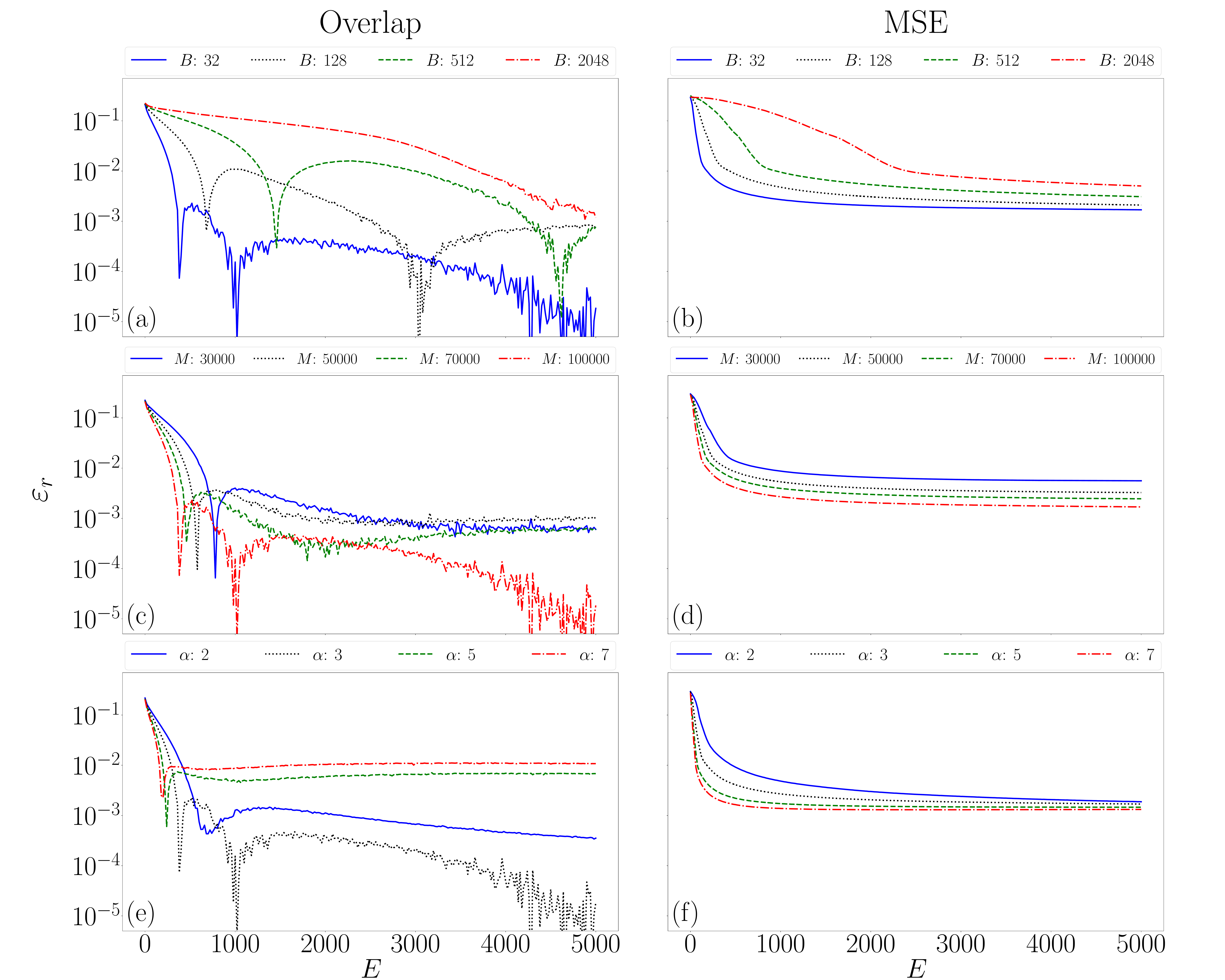} 
    \caption{Relative error $\nrgerr$ for the ferromagnetic ground state ($J_1=1$ and $J_2=2$) versus the number of epochs $E$. In panels (a, c, e), we use overlap loss function $\lossover$; in panels (b, d, f), we use $\lossmse$. In panels (a,b), we vary the batch size $B$. In panels (c,d), we vary the number of samples $M$; in panels (e,f), we vary the hidden ratio $\alpha$. If not stated otherwise, the number of training samples is $M=10^5$, the hidden ratio is $\alpha=3$, the batch size is $B=32$ and $\gamma=10^{-5}$ is the initial learning rate.}
    \label{fig:ferro}
\end{figure}

We consider the ferromagnetic phase, for Hamiltonian parameters $J_1 = 1$ and $J_2 = 2$. The default values, unless otherwise stated, are $M = 10^5$ for the number of samples, $\alpha=3$ for the hidden ratio, $B=32$ for the batch size, and $\gamma=10^{-5}$ for the initial learning rate. The results are shown in Figure~\ref{fig:ferro}, in a similar structure as in Figure~\ref{fig:para}.  

In Figure~\ref{fig:ferro}, we observe that in some cases, the error becomes very small, and then it increases. This is due to the fact that in those cases, the energy $E_T$ obtained from training the neural network starts from below the ground state value, then goes above it, and then may decrease again when approaching the exact value from above. 
We point out that this is not at odds with the variational principle, for which the energy is bounded from below compared to that of the ground state. This is because, when sampling, we may not evaluate the Hamiltonian correctly and it is thus possible to estimate an energy value lower than the exact ground state energy. 
Overall, in the ferromagnetic phase, we observe a similar behavior as with the paramagnetic phase. However, we can reach more accurate predictions of the energy with a relative error of the order of $10^{-3}$ or even smaller. 
Smaller batch sizes are generally helpful for both loss functions, and so are more number of samples, while a larger number of hidden nodes seems to lead to overfitting for $\lossover$, Figure~\ref{fig:ferro}(e), while it still allows (marginal) improvements for $\lossmse$, Figure~\ref{fig:ferro}(f). 
We associate this overfitting to the relative simplicity of the wave function to be described by the network. 
We still observe that $\lossover$ allows the energies to reach values closer to the ground state energy. 

\subsection{Training in the pair-antiferromagnetic phase}

\begin{figure}[ht]
\centering
\includegraphics[width=0.95\columnwidth]{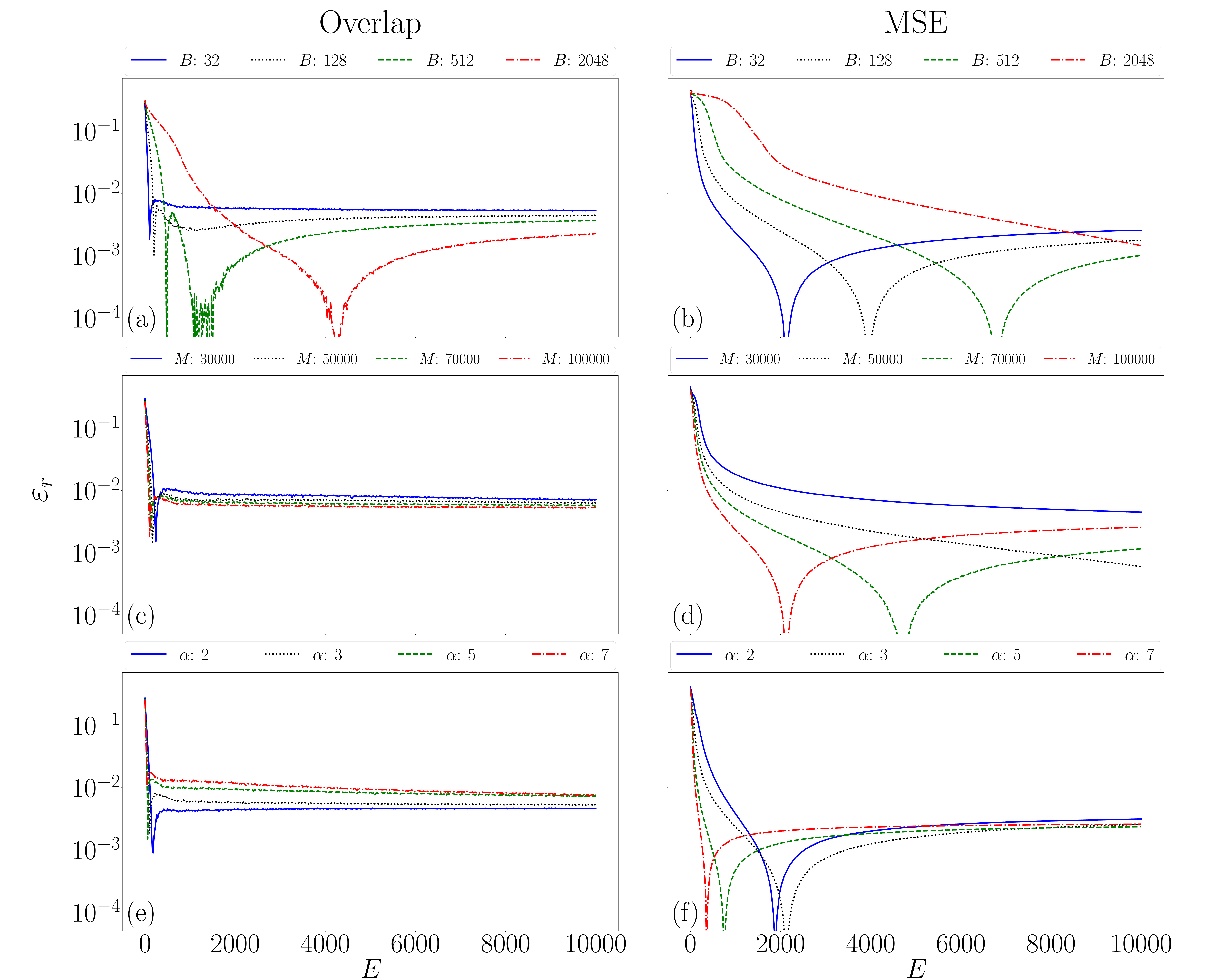} 
    \caption{Relative error $\nrgerr$ for the pair-antiferromagnetic ground state ($J_1=1$ and $J_2=-2$) versus the number of epochs $E$. In panels (a, c, e), we use overlap loss function $\lossover$; in panels (b, d, f), we use $\lossmse$. In panels (a,b), we vary the batch size $B$. We vary the number of samples $M$ in panels (c,d); in panels (e,f), we vary the hidden ratio $\alpha$. If not stated otherwise, the number of training samples is $M=10^5$, the hidden ratio is $\alpha=3$, the batch size is $B=32$ and $\gamma=10^{-5}$ is the initial learning rate.}
    \label{fig:pairantiferro}
\end{figure}

In the pair-antiferromagnetic phase, for $J_1=1$ and $J_2=-2$, we presented the results in Figure~\ref{fig:pairantiferro} following the same structure as Figure~\ref{fig:para} and Figure~\ref{fig:ferro}. The default values of the training parameters are $M = 10^5$ for the number of samples, $\alpha=3$ for the hidden ratio, $B=32$ for the batch size, and $\gamma=10^{-5}$ for the initial learning rate.  
The shown curves result from the fact that the initial energy of the state is below the ground state energy, and that it requires a larger number of epochs to approach the ground state. We computed $10^4$ epochs here as the model has slower optimization dynamics. In general, we still observe some form of overfitting for $\lossover$, Figure~\ref{fig:pairantiferro}(e), and a general improvement of performance with an increasing number of samples $M$, Figure~\ref{fig:pairantiferro}(c,d). As for the batch size, a smaller batch size results in approaching the ground state energy in a smaller number of epochs, see Figure~\ref{fig:pairantiferro}(a,b).

\subsection{Convergence of loss function and relative energy error} 

\begin{figure}[ht]
\centering
\includegraphics[width=\columnwidth]{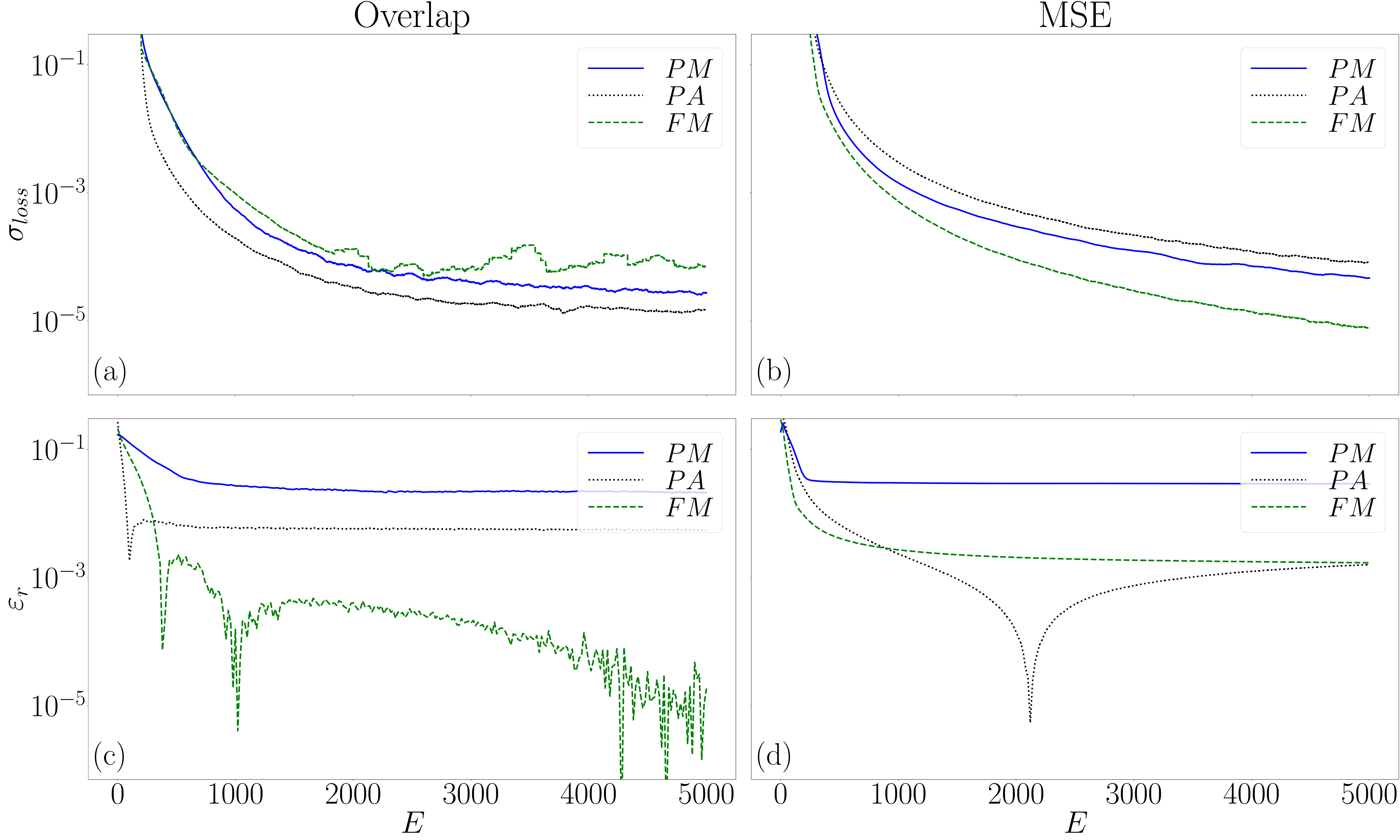} 
    \caption{(a,b) Variance of the loss function value $\sigma_{loss}$ verus number of epochs $E$. (c,d) Relative energy error $\nrgerr$ versus the number of epochs $E$. Panels (a,c) are for $\lossover$, and panels (b,d) are for $\lossmse$. The blue continuous line is for the paramagnetic (PM) phase  $J_1=2$ and $J_2 = -1$, the grey dotted line for the pair-antiferromagnetic (PA) phase  $J_1 = 1$ and $J_2 = -2$, and the green dashed line for the ferromagnetic  (FM) phase $J_1 = 1$ and $J_2 = 2$. In all panels we have used $B=32$, $M=10^5$, $\alpha=3$ and $\gamma=10^{-5}$. }      
    \label{fig:loss-std-with-eloc} 
\end{figure}

In supervised training one typically only has access to data such as configurations and their corresponding probability amplitude, and not to the Hamiltonian of the system. During the training, we can monitor the evolution of the loss function, but this does not directly imply that other properties of the system, such as the energy, are better represented. We thus study the convergence of a loss function and compare that to the relative energy error $\nrgerr$. For the loss function, we evaluate the variance of the loss function $\sigma_{loss}$ over the past 200 epochs and we plot it against the epoch number. This is because a natural stopping criterion for the optimization is indeed that the variation of the loss function over the epochs is below a certain threshold.

In Figure~\ref{fig:loss-std-with-eloc} we show this comparison between the standard deviation of the loss function $\sigma_{loss}$ over the past 200 epochs, panels (a,b), and the relative energy error $\nrgerr$ panels (c,d), both versus the number of epochs $E$.  
In panels (a,c) we consider the training with the loss function $\lossover$, while in panels (b,d) we used the loss function $\lossmse$. In each panel the blue continuous line corresponds to the paramagnetic phase ($J_1 = 2$ and $J_2 = -1$), the grey dotted line to the pair-antiferromagnetic phase  ($J_1 = 1$ and $J_2 = -2$), and the green dashed line to the ferromagnetic phase ($J_1 = 1$ and $J_2 = 2$). 
We observe that, generally, a decrease in the loss function indeed results in a wave function with energy closer to the ground state energy, however, a quantitative correspondence between the two is not guaranteed.

\section{Conclusion}\label{sec:conclusions}

We have studied the supervised training of a neural network to represent the ground state of a many-body quantum system. We considered systems, and their corresponding Hamiltonians, in various phases of matter, namely, the paramagnetic phase, the ferromagnetic phase, and the pair-antiferromagnetic phase. We considered two different loss functions for the training of the neural network, one that tries to maximize the overlap between the trained data and the neural network wave function, $\lossover$, and one that tries to reduce the difference between the logarithm of the probability amplitudes and the output of the neural network $\lossmse$. 

As possible strategies for supervised learning, we observed that the loss function $\lossover$ generally results in wave functions with more accurate ground state energy in all the phases considered.
We associate this performance to the fact that the overlap loss function does not require an exact match of the predicted output with the training data, but instead it can be proportional. This would be relevant also for other types of Hamiltonians.
However, it is important that for overlap, the neural network output is rescaled to increase the stability of the algorithm, see Ref.~\cite{VicentiniCarleo2021} and~\ref{app:notrescaled}. In our analysis of the hyperparameters, we found that smaller batch sizes result in better performance, but this also requires a larger number of epochs to converge (see~\ref{app:batch}). Furthermore, larger sample sizes generally help to obtain better results, but larger hidden dimensions can result, especially when using $\lossover$, in overfitting when dealing with relatively simple wave functions. 

Further directions include investigating the supervised learning of wave functions that may not be described by real positive numbers, for example, for a more complex ground state, an excited state (but still an eigenstate of the system), or even non-equilibrium states which behave like steady states~\cite{XuPoletti2022, XuPoletti2023}. 



\section{Acknowledgments}\label{Acks}
S.B. acknowledges support from the Ministry of Education Singapore, under the grants MOE-T2EP50120-0019 and MOE- T1 251RES2302. D.P. acknowledges support from the Ministry of Education Singapore under the grant MOE-T2EP50120-0019 and the National Research Foundation, Singapore under its QEP2.0 programme (NRF2021-QEP2-02-P03). 
The computational work for this article was partially performed at the National Supercomputing Centre, Singapore \cite{NSCC}. 
The codes and data generated are both available upon reasonable request to the authors.



\bibliographystyle{elsarticle-num} 
\bibliography{Bibliography.bib}
\biboptions{sort&compress,square}


\clearpage 
\newpage
\appendix 

\setcounter{figure}{0} 
\renewcommand{\thefigure}{A\arabic{figure}}

\section{Not rescaled wave function for overlap KL divergence loss functions}\label{app:notrescaled}    

\begin{figure}[h]
\includegraphics[width=\columnwidth]{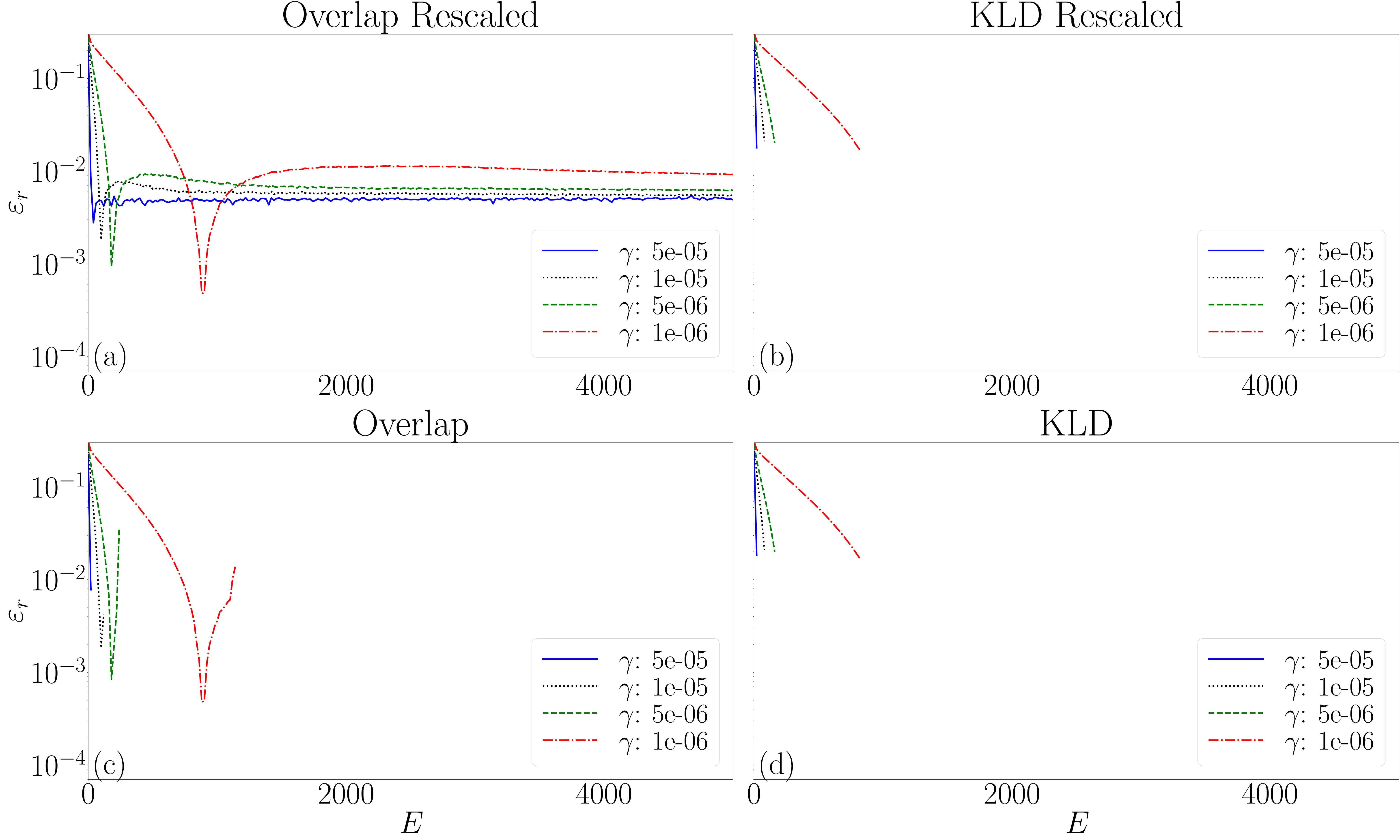} 
    \caption{Relative error $\nrgerr$ for the pair-antiferromagnetic ground state ($J_1=1$ and $J_2=-2$) versus number of epochs $E$. (a) is the rescaled version of overlap. (b) is the rescaled version of KL divergence. (c) is the original overlap and (d) is the original KL divergence. We vary to different learning rates compared to the control/default value $\gamma=10^{-5}$ in the paper. the number of training sample is $M=10^5$, batch size $B=32$, and the hidden ratio is $\alpha=3$.} 
    \label{fig:exploding-pa}
\end{figure} 

\begin{figure}[h]
\vspace{0.5cm}
\includegraphics[width=\columnwidth]{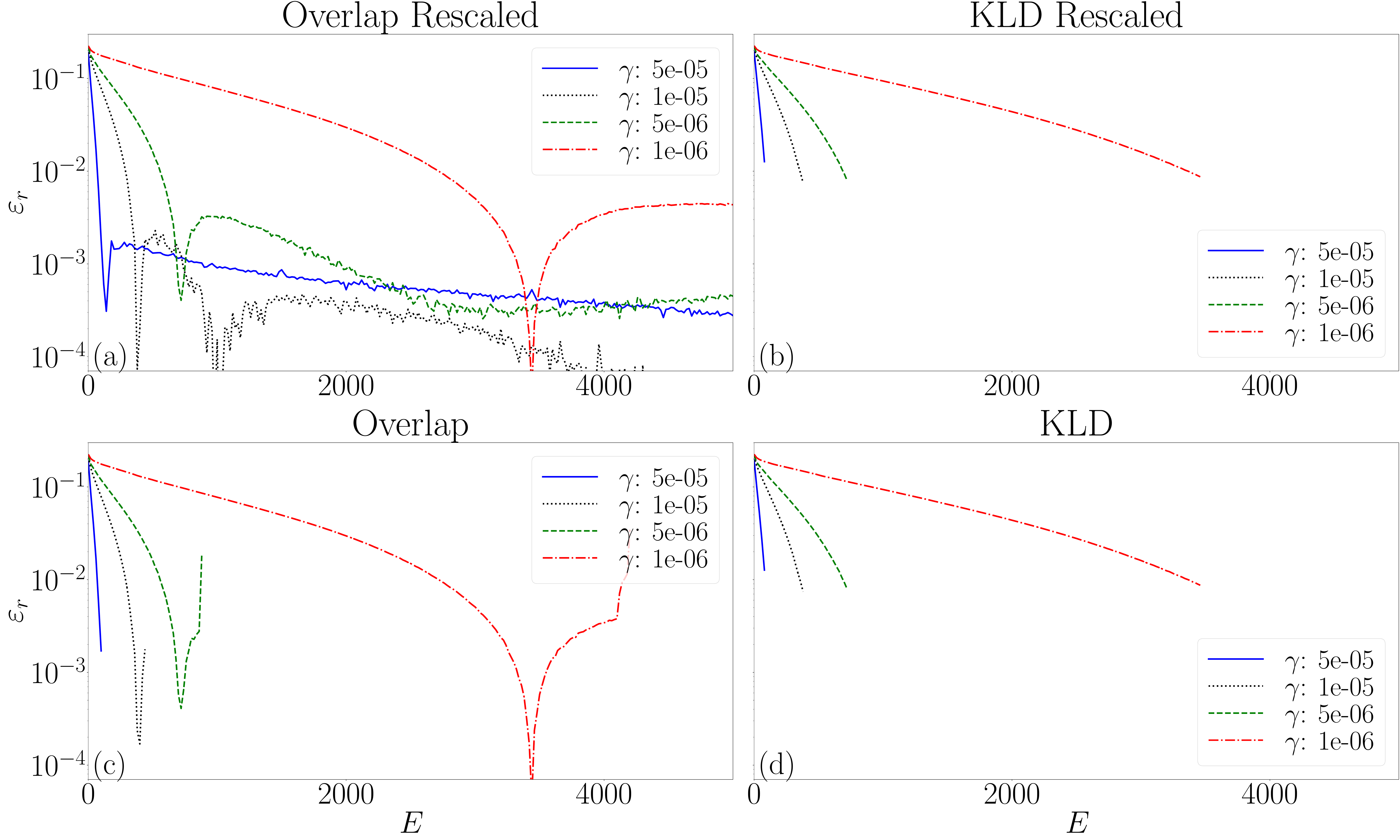} 
    \caption{Relative error $\nrgerr$ for the ferromagnetic ground state ($J_1=1$ and $J_2=2$) versus number of epochs $E$. (a) is the rescaled version of overlap. (b) is the rescaled version of KL divergence. (c) is the original overlap and (d) is the original KL divergence We vary to different learning rates compared to the control/default value $\gamma=10^{-5}$ in the paper. the number of training sample is $M=10^5$, batch size $B=32$, and the hidden ratio is $\alpha=3$.}
    \label{fig:exploding-fm}
\end{figure} 

In the main portion of the article, when using $\lossover$, we rescale the neural network output using Equation~(\ref{NNoutput}). Here we show that if we were not to do this, the training would be more unstable and it could lead to the network parameters becoming ${\rm NaN}$. This is a general result, as we show by discussing two loss functions, the overlap loss function $\lossover$ and the Kullback–Leibler (KL) divergence. In this work, the loss function for KL divergence is given by Equation~(\ref{KLloss}).
\begin{align}
    \mathcal{L}_{\rm KLD}(\Psi_G, \Psi) = \sum_{m=1}^{M_u} {|\Psi_G( {\bm x}_m )|^2}\ln\left( \frac{|\Psi_{G}( {\bm x}_m )|^2}{|\Psi( {\bm x}_m )|^2} \right)\label{KLloss}
\end{align}
where the wave functions could be rescaled, i.e. each one is divided by the maximum value obtained over the $M_u$ unique samples, or not. 
Our results are shown in Figure~\ref{fig:exploding-pa} and Figure~\ref{fig:exploding-fm}, respectively for parameters such that the ground state is in the pair-antiferromagnetic and in the ferromagnetic phase (the paramagnetic phase is more stable). In the figures, an interrupted line before 5000 epochs signifies that the training leads to ${\rm NaN}$ for the network parameters. Importantly, we note that even using a rescaled neural network for the KL divergence loss function can result in instability of the learning dynamics, as shown by the interruption of the lines. We associate the instability of the KL divergence loss function to the highly peaked distributions that we are using for the training.

In general, however, we note that using a smaller initial learning rate $\gamma$ results in a delayed instability of the neural network parameters' dynamics.

\section{Smaller batch size for training}\label{app:batch} 
\begin{figure} 
\vspace{0.5cm} 
\centering
\includegraphics[width=0.55\columnwidth]{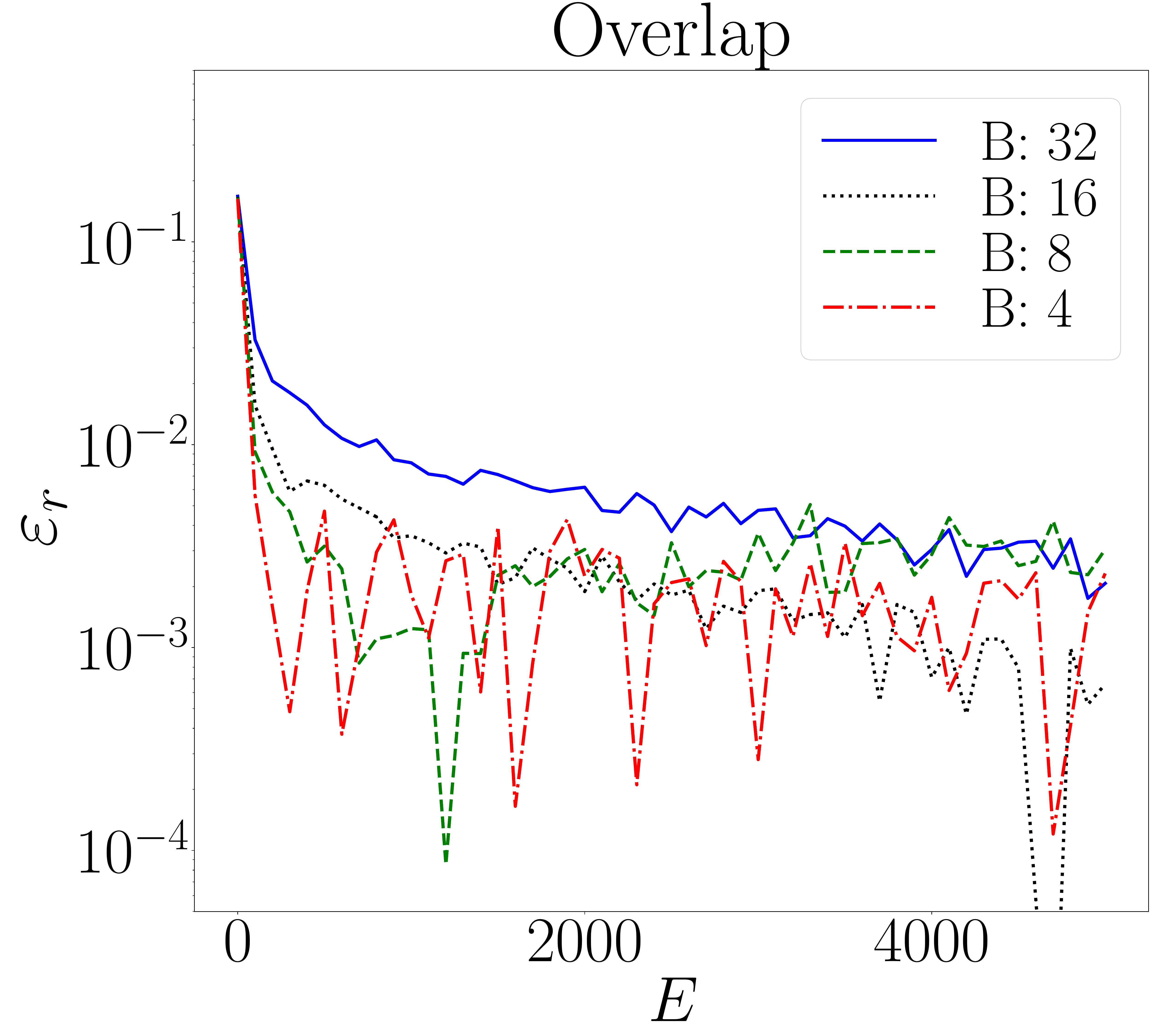} 
    \caption{Relative error $\nrgerr$ for the paramagnetic ground state ($J_1=2$ and $J_2=-1$) versus number of epochs $E$. we use overlap loss function $\lossover$ and vary to smaller batch size compared to the control/default value $B=32$ in the paper. The number of the training samples is $M=10^5$, the learning rate is $10^{-4}$ and the hidden ratio is $\alpha=3$. } 
    \label{fig:batch} 
\end{figure} 

In the main portion of the article, we have observed that the smaller the batch size the better the wave function approaches the ground state energy. Here we explore this further with even smaller batch sizes using $\lossover$ as a loss function. 
In Figure~\ref{fig:batch} we show results for 128 spins in the  paramagnetic phase versus the number of epochs. We observe that smaller batch sizes can result in larger oscillations and not significant improvements while, not shown, requiring more time. For these reasons, we kept $B=32$ as a standard minimum value for the computations in the main portion of the article.  We highlight here that for Figure~\ref{fig:batch} we have used an initial learning rate $\gamma=10^{-4}$ which, for $\lossover$ and in the paramagnetic phase, allows us to reach energies closer to the ground state  than using $\gamma = 10^{-5}$. The overall dependence on the batch size $B$ is qualitatively similar for the two values of learning rate $\gamma$.

\section{About the \textsc{Mapalus}}\label{app:mapalus}
In \textsc{Mapalus},  users can use neural-network quantum states to find the ground state or excited states of a given system. The system is described by the structure and the quantum model of the system. The library non-exclusively readily contains the implementation for the transverse field Ising model, the Heisenberg XYZ model, and the Heisenberg J1-J2 model.  Users can choose to train a neural-network quantum states in either an unsupervised or a supervised manner with different neural network architectures (the library currently contains the code for multilayer feed forward neural network, restricted Boltzmann machines, and deep Boltzmann machines) with dynamic and static stopping criteria, and different Monte Carlo sampling algorithms. The library also provides a logging mechanism that saves the necessary information from the training. The logging process also includes the computation of observables. \textsc{Mapalus} is written in Python with \textsc{TensorFlow}. It is easily customizable and runs in parallel on Message Passing Interface or general-purpose graphics processing units.

Using \textsc{Mapalus}, we only needed to implement the different loss functions considered in order to empirically explore the respective effectiveness of the different configurations and parameters' values of the supervised multi-layer  feed forward neural network quantum states.

\end{document}